\documentclass[%
 reprint,
 superscriptaddress,
 nofootinbib,
 amsmath,amssymb,
 aps,
 prl,
]{revtex4-2}

\usepackage{graphicx}% Include figure files
\usepackage{dcolumn}% Align table columns on decimal point
\usepackage{bm}% bold math
\usepackage{booktabs}
\usepackage[english]{babel}
\usepackage{amsmath,amssymb,amsbsy,amstext, amsthm, simplewick}
\usepackage{hyperref}
\usepackage{graphicx}
\usepackage{amsfonts}
\usepackage{amssymb}
\usepackage{upgreek}
\usepackage{cancel}
\usepackage{color}
\usepackage{slashed}
\usepackage[dvipsnames]{xcolor}
\usepackage{tikz-cd}
\usepackage{feynmp-auto}
\usepackage{color}
\usepackage{soul}
\usepackage{dsfont}
\usepackage{verbatim}
\usepackage[utf8]{inputenc} 
\usepackage{soul}

\usepackage{empheq}
\newlength\dlf  % Define a new measure, dlf

\def\@fnsymbol#1{\ensuremath{\ifcase#1\or $\Re$\or $\Im$\or  \else\@ctrerr\fi}}

\begin{document}

%\preprint{APS/123-QED}

\title{Hydrogen Mixing as a Novel Mechanism for Colder Baryons in 21 cm Cosmology}% Force line breaks with \\

\author{Lucas Johns}
\email{NASA Einstein Fellow (ljohns@berkeley.edu)}
\affiliation{Departments of Astronomy and Physics, University of California,
Berkeley, CA 94720, U.S.A.}
\affiliation{Department of Physics, University of California, San Diego, CA 92093, U.S.A.}
\author{Seth Koren}
\email{EFI Oehme Fellow (sethk@uchicago.edu)}
\affiliation{Enrico Fermi Institute, University of Chicago,
Chicago, IL 60637, U.S.A.}
\affiliation{Department of Physics, University of California,
Santa Barbara, CA 93106, U.S.A.}


\begin{abstract}
The anomalous 21 cm absorption feature reported by EDGES has galvanized the study of scenarios in which dark matter (DM) siphons off thermal energy from the Standard Model (SM) gas. In a departure from the much-discussed models that achieve cooling by DM scattering directly with SM particles, we show that the same end can be achieved through neutral atomic hydrogen $H$ mixing with a degenerate dark sector state $H'$. An analysis of in-medium $H$--$H'$ oscillations reveals viable parameter space for generic types of $H'$--DM interactions to provide the requisite cooling. This strategy stands in stark contrast to other proposals in many respects, including its cosmological dynamics, model building implications, and complementary observational signatures.
\end{abstract}

\maketitle

%\tableofcontents

Before the first stars cast their light, the cosmos was a dim but dynamic place. The long period known as the cosmic dark ages saw the formation of halos and galaxies as the universe sowed the seeds of its later illumination. But while dark, the era was not pitch black. Photons from the cosmic microwave background (CMB) streamed everywhere, interacting with the hyperfine transition of hydrogen and reaching us, today, as messengers.

In 2018 the Experiment to Detect the Global Epoch of Reionization Signature (EDGES) announced a detection of 21 cm hyperfine absorption at redshifts $15 \lesssim z \lesssim 20$ \cite{Bowman:2018yin}. Their signal, though contested \cite{2018Natur.564E..32H,2018Natur.564E..35B,2019ApJ...874..153B,2019ApJ...880...26S,2020ApJ...897..132T,2019MNRAS.489.4007S,2020MNRAS.492...22S}, may indicate that the hydrogen gas at cosmic dawn was anomalously cold, at the level of $3.8 \sigma$ \cite{Barkana:2018qrx,Barkana:2018lgd}. If this is correct, the EDGES measurement demands a reassessment of the physics of the dark ages.

An enticing explanation is that the gas lost some of its thermal energy to dark matter (DM) in the lead-up to cosmic dawn \cite{Bowman:2018yin,Barkana:2018lgd}. In this paper we adopt that interpretation and develop the \textit{hydrogen mixing portal}---mixing between neutral atomic hydrogen $H$ and a dark sector state $H'$---as a new mechanism for transferring heat from the Standard Model (SM) sector to the dark one.

The mechanism stands in contrast to the well-known proposal that the cooling was facilitated by scattering between DM and SM particles \cite{Barkana:2018qrx,Barkana:2018lgd,Munoz:2018pzp,Jia:2018csj,Berlin:2018sjs,Slatyer:2018aqg,Fraser:2018acy,Kovetz:2018zan,Liu:2019knx}. Although models of the latter type are tenable, the most elegant of them are strongly constrained by fifth force experiments, cosmology, and other observations. In thinking about how else the gas might be cooled by new particle physics, we are guided by the hint that the EDGES anomaly appears in connection with \textit{neutral} hydrogen. One of the special attributes of a neutral particle is that it can oscillate into a dark state without violating SM gauge symmetry. In this sense, $H$--$H'$ mixing is in kinship with, for example, the neutrino portal (between SM and sterile neutrinos) and the kinetic mixing portal (between SM and dark photons).

To get a sense of the relevant energy scale of mixing, we make note of the following. For a hydrogen atom in the late dark ages to undergo $\sim 1$ period of oscillation in between scattering events, the mixing parameter $\delta$ must be comparable to the hydrogen scattering rate $\Gamma_H$ at the relevant redshifts. For $z = 30$, this condition translates to $\delta \sim \mathcal{O}(10^{-37})$~GeV.

It is immediately apparent that $H$--$H'$ mixing is exquisitely sensitive to any physics that distinguishes the two states. This means that hydrogen oscillations are not likely to show up in any terrestrial experiment---interactions easily swamp the mixing term---but it also indicates that even a tiny mass splitting can throttle the hydrogen portal. We therefore assume that $H$ and $H'$ are exactly degenerate. In a companion paper \cite{companion} we discuss $H'$ as \textit{mirror hydrogen}, a composite particle related to $H$ by an exact $\mathbb{Z}_2$ symmetry. While we mention that example a couple times below, we note that all the cosmology we discuss herein is independent of the identity of $H'$ past its mass degeneracy with $H$.

How exactly does mixing bring about cooling of the gas? For a single atom, the sequence can be thought of as occurring in steps:
\begin{equation}
H(p) \xrightarrow{\textrm{osc}} H'(p) \xrightarrow{\textrm{$H'$--$X$}} H'(p') \xrightarrow{\textrm{osc}} H(p').
\end{equation}
A SM hydrogen atom with momentum $p$ oscillates into mirror hydrogen, scatters with DM (which we call $X$), and oscillates back into SM hydrogen, now possessing a smaller momentum $p'$.

\begin{figure}
    \centering
    \includegraphics[width=0.45\textwidth]{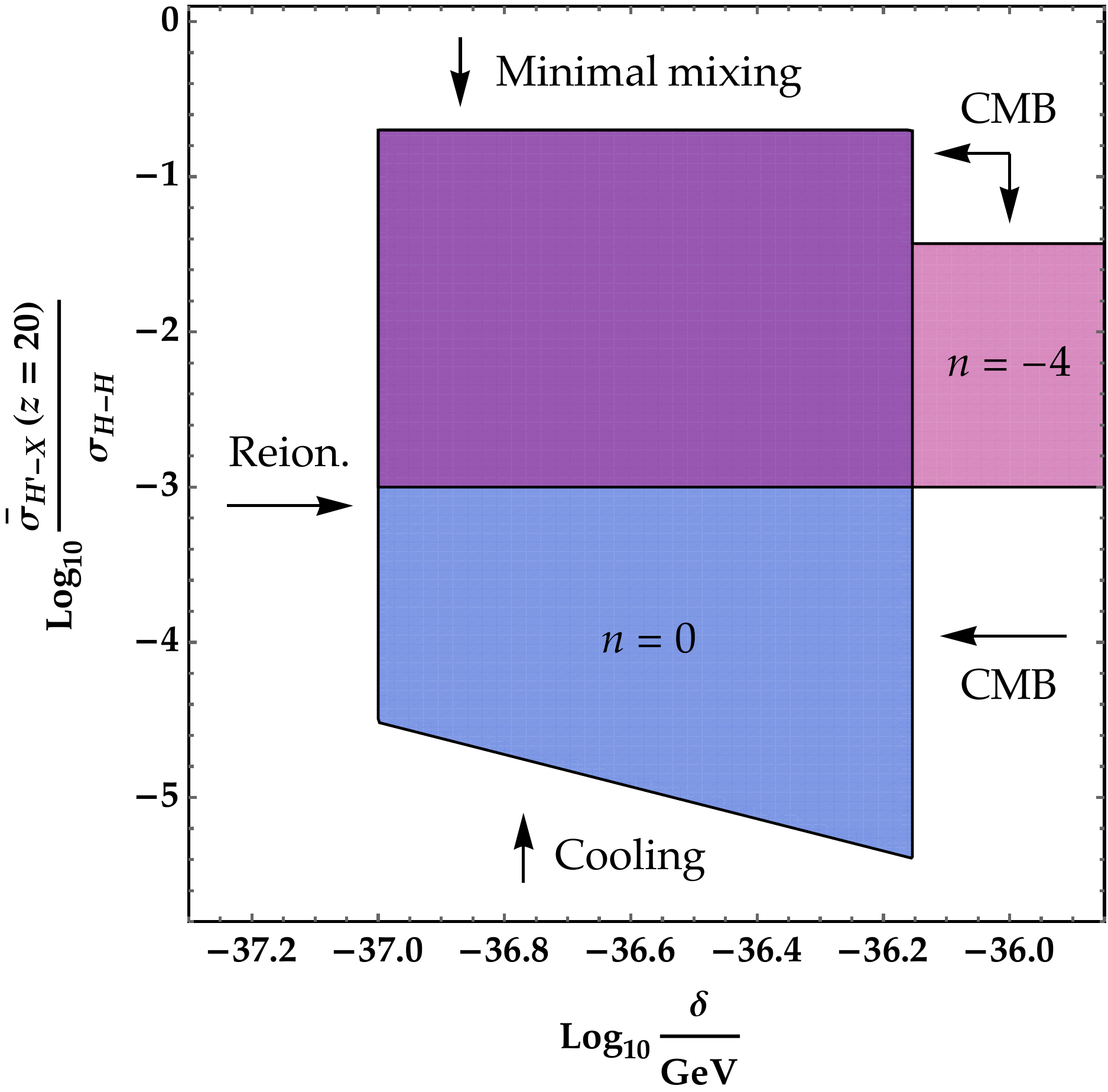}
    \caption{Viable parameter space for cooling through $H$--$H'$ mixing, adopting $H'$--$X$ cross section $\bar{\sigma}_{H'-X} = \sigma_0 v_m^n$ with $n=0$ (blue) and $n=-4$ (pink). The cross section is expressed as a fraction of $\sigma_{H-H} = 4 \pi a_0^2$. Consistency with EDGES (\textit{Cooling}) implies lower limits on $\bar{\sigma}_{H'-X}$ for both cases shown. Indirect $H$--$X$ coupling at $z \gtrsim 200$ (\textit{CMB}) generally bounds $\delta$ from above but leaves open a swath for $n=-4$. Other labeled regions are excluded for reasons described in the main text. This analysis assumes that $H'$ and $X$ thermally equilibrate, from which it follows that $m_X \sim 2$~GeV.}
    \label{fig:minmax}
\end{figure}

Figure~\ref{fig:minmax} shows a conservative parameter space in which this sequence thermally equilibrates $H'$ and $X$ and cools the gas to the temperature inferred from the EDGES measurement. The free parameters are $\delta$ and the momentum-transfer cross section $\bar{\sigma}_{H'-X}$. Text labels in Fig.~\ref{fig:minmax} summarize the major considerations that are explicated below. We observe that the viable range in $\delta$ is indeed at roughly the scale estimated a few paragraphs up. While a careful analysis of the mixing dynamics is of course more nuanced than the simple estimate, their agreement indicates that the preferred region of $\delta$ is picked out fairly straightforwardly by cosmology. Coincidentally, the region is also at an interesting scale for high energy physics. In the case where $H'$ is identified with mirror hydrogen in a $\mathbb{Z}_2$-symmetric model, we may estimate \cite{companion}
\begin{equation}
\delta \sim \frac{(4 \pi)^2 \Lambda_\textrm{QCD}^6}{\Lambda^8}\frac{1}{a_0^3}, \label{lambda}
\end{equation}
where $\Lambda$ is the ultraviolet scale at which the hydrogen-mixing interactions are generated and the factors of the Bohr radius $a_0$ arise from the wavefunction overlap between the electron and proton. One finds automatically that $\delta \sim 10^{-37}$~GeV arises from new physics appearing at a scale $\Lambda \sim 100~\textrm{GeV} - 1$~TeV, in sight of the energy frontier. The details of the mirror matter model implementation of this mechanism are explored in the companion paper \cite{companion}, in addition to other connections to particle physics.

\begin{figure*}
    \centering
    \includegraphics[width=0.8\textwidth]{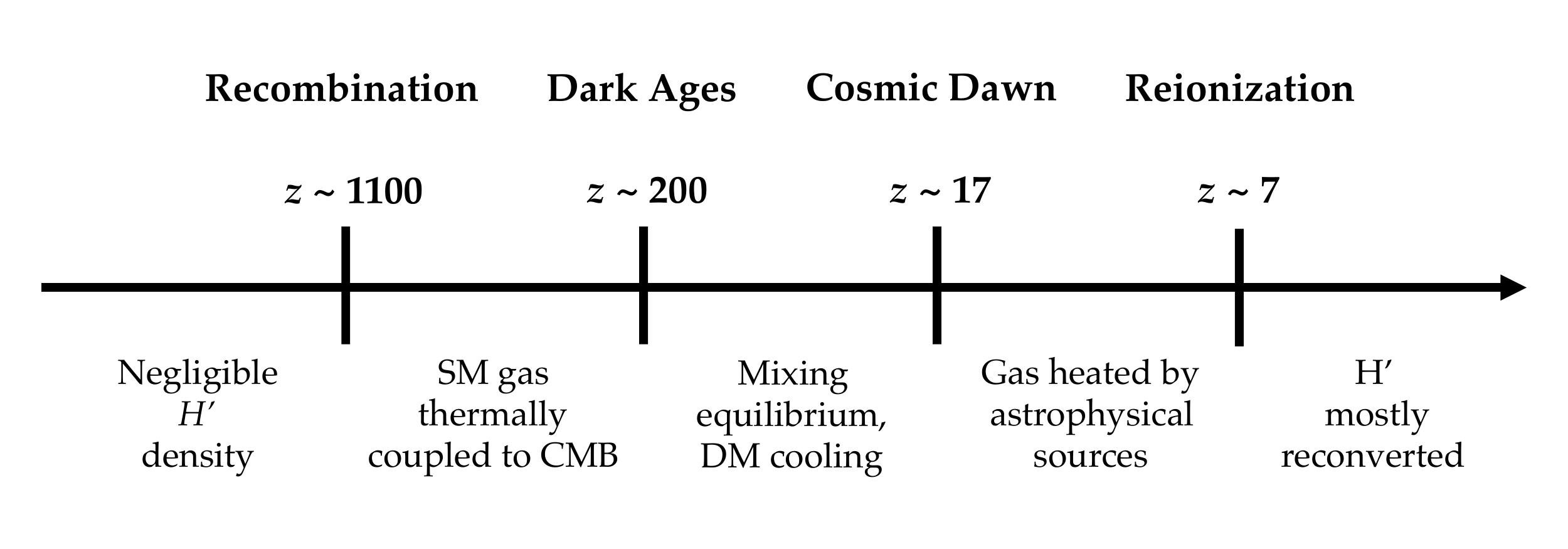}
    \caption{Cosmic timeline schematically marking the significant epochs in the proposed scenario. Sometime between $20 \lesssim z \lesssim 200$, $H$--$H'$ mixing comes into (or continues to be in) equilibrium and both species are cooled by interactions between $H'$ and dark matter (DM). Mixing equilibrium is maintained during the epoch of reionization, and the $H'$ population depletes along with the neutral abundance of SM hydrogen.}
    \label{fig:timeline}
\end{figure*}

We now describe the hydrogen portal scenario in greater detail. Figure~\ref{fig:timeline} orients the mixing dynamics with respect to certain key events and phases in the cosmic timeline. Prior to recombination ($z \sim 1100$), the $H'$ density is vanishingly small. Oscillations begin once neutral atoms form and the universe enters the dark ages, but efficient cooling through $H'$--$X$ scattering must wait at least until $z \sim 200$, after which the hydrogen gas is no longer thermally coupled to the CMB. The net effect of oscillations and scattering-induced decoherence is to chemically equilibrate $H$ and $H'$. With the hydrogen species in chemical (\textit{i.e.}, mixing) equilibrium and $H'$ and $X$ in thermal equilibrium, the SM hydrogen density dilutes by half ($n_H = n_{H'}$) and all three temperatures reach a common value ($T_H = T_{H'} = T_X$). We restrict our analysis to the regime where this is fully attained by cosmic dawn. 

The timeline also marks $z \sim 17$, the redshift at which the EDGES signal reaches its maximum amplitude and astrophysical sources begin to significantly heat the gas. At this stage, the SM sector still only contains half the baryons it had at recombination. It would be flagrantly unacceptable for this to remain true down to the present day: the cosmic baryon density $\eta$ is known to high precision, and more than 50\% of it is accounted for by low-$z$ observations \cite{2018Natur.558..406N,deGraaff2019,Johnson_2019,Kov_cs_2019,Macquart:2020lln}. A large part of the $H'$ population that is produced must ultimately return to our own sector if the scenario is not to run afoul of the low-$z$ baryon tally. This reconversion occurs naturally if mixing is efficient during the epoch of reionization ($z \sim 7$), as oscillations keep the species in chemical equilibrium, causing $n_{H'}$ to track $n_H$ (the \textit{neutral} density) as it declines.

Other versions of the timeline are conceivable---DM might never fully thermalize with $H'$, for example---but we focus only on the one just described, which is the simplest of them. 

The $H$--$H'$ density matrix $\rho$ evolves according to the quantum kinetic equation
\begin{equation}
i \frac{d\rho}{dt} = \left[ \mathcal{H}, \rho \right] + i\mathcal{C}, \label{qke}
\end{equation}
where $\mathcal{H}$ is the Hamiltonian and $\mathcal{C}$ is the collision term. Since $H$ and $H'$ are degenerate in vacuum, the Hamiltonian is
\begin{equation}
\mathcal{H} = \begin{pmatrix} \Delta V & \delta \\ \delta & - \Delta V \end{pmatrix}. \label{oscham}
\end{equation}
The potential $\Delta V = (V_H - V_{H'} )/2$ stems from forward scattering in the medium. Hydrogen of wave number $k$ experiences an index of refraction
\begin{equation}
n = 1 + \sum_i \frac{2 \pi n_i}{k^2} f_i (0),
\end{equation}
where $n_i$ and $f_i (0)$ are the target number density and elastic forward scattering amplitude associated with process $i$.

We parametrize the mixing in terms of an in-medium oscillation frequency $\omega_m = ( \Delta V^2 + \delta^2 )^{1/2}$ and an in-medium mixing angle given by $\sin^2 2 \theta_m = \delta^2 / \omega_m^2$. Equilibration of the mixing channel then occurs on a timescale $\Gamma_\textrm{osc}^{-1}$, with
\begin{equation}
\Gamma_\textrm{osc} \sim \frac{\Gamma_c}{4} \frac{\sin^2 2 \theta_m}{1 + \left( \Gamma_c / 2 \omega_m \right)^2}, \label{gosc}
\end{equation}
where $\Gamma_c = \Gamma_H + \Gamma_{H'}$ is the sum of the $H$ and $H'$ scattering rates. Eq.~\eqref{gosc} is often used in the literature on sterile neutrino dark matter, going back to Refs.~\cite{KAINULAINEN1990191,PhysRevLett.68.3137,PhysRevLett.72.17} and others. It follows quite immediately from applying a relaxation-time approximation to the evolution of $\rho$ \cite{PhysRevD.100.083536}.

During the dark ages, the most important interaction for mixing is $H$--$H$ scattering. It gives a good approximation of the total collisional rate of hydrogen:
\begin{equation}
\Gamma_H \sim n_H \left\langle \sigma v \right\rangle_{H-H} \sim \left( 1 \times 10^{-43}~\textrm{GeV} \right) \left( 1+z \right)^4. \label{gammaheq}
\end{equation}
The same is true of the forward-scattering potential:
\begin{equation}
V_H \sim \frac{2 \pi n_H a_0}{m_H} \sim \left( 3 \times 10^{-42}~\textrm{GeV} \right) \left( 1+z \right)^3. \label{vheq}
\end{equation}
In the equations above, $\langle \sigma v \rangle$ is the velocity-averaged $H$--$H$ cross section and $a_0$ is the Bohr radius. Contributions from other processes are considered in Ref.~\cite{companion} and shown to be subdominant. For the sake of completeness, we include them in the results presented here.

\begin{figure}
    \centering
    \includegraphics[width=0.45\textwidth]{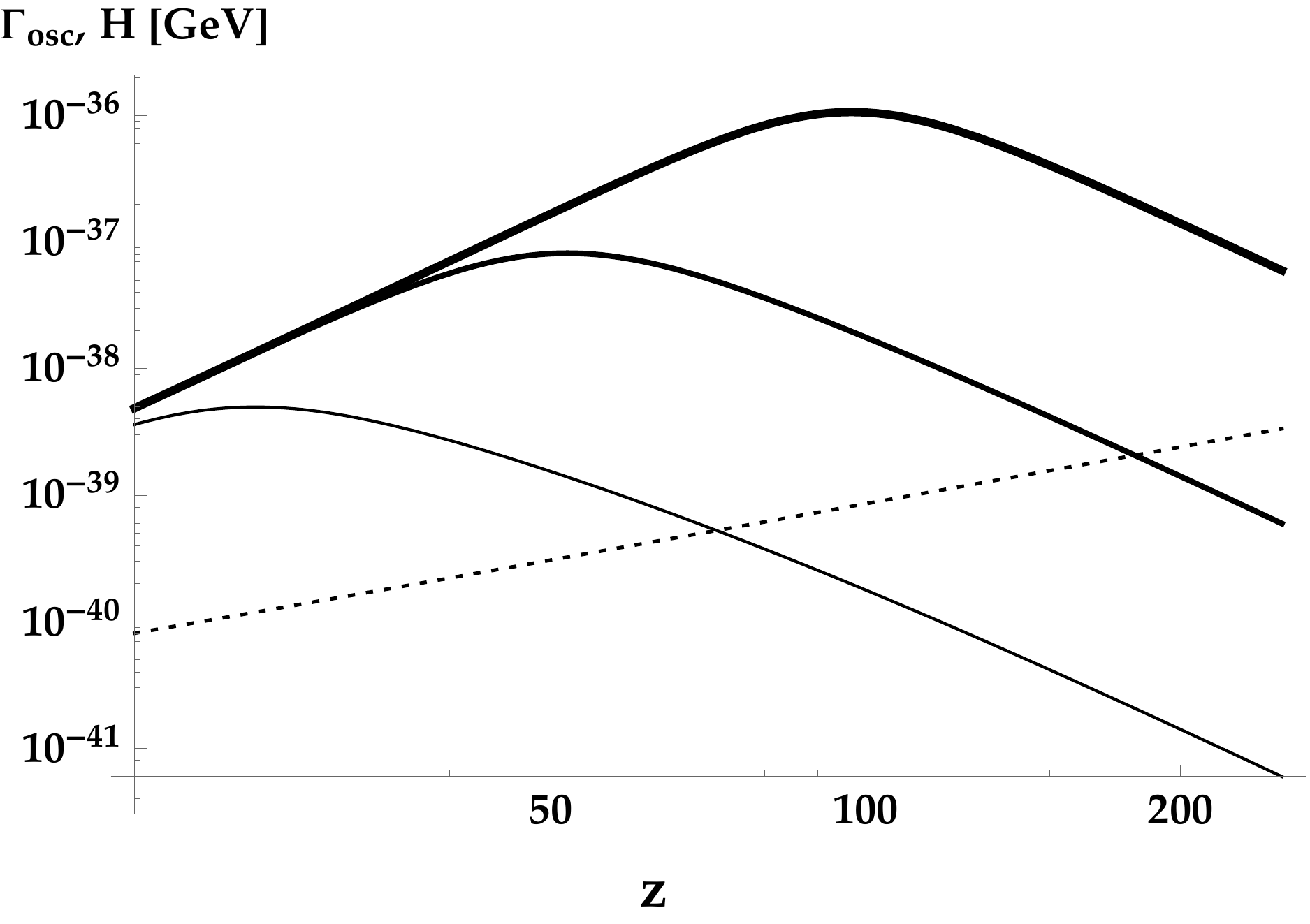}
    \caption{Hubble rate $H$ (dotted) compared with the mixing equilibration rate $\Gamma_\textrm{osc}$ at $\delta = 5 \times 10^{-38}$~GeV (thin), $5 \times 10^{-37}$~GeV (medium), and $5 \times 10^{-36}$~GeV (thick). $\Gamma_\textrm{osc}$ captures the effects of forward scattering and quantum decoherence on vacuum oscillations. The dominant interaction is $H$--$H$ scattering.}
    \label{fig:gosc}
\end{figure}

Figure~\ref{fig:gosc} presents the evolution over redshift of $\Gamma_\textrm{osc}$ for three different values of $\delta$. Feedback on the mixing from $H'$--$H'$ scattering---an unimportant effect for the purposes of estimating timescales---is not included. More significantly, the figure assumes a scenario of \textit{minimal mixing}, by which we mean that $H'$--$X$ scattering does not substantially change the mixing dynamics. Although this stipulation is not a constraint per se, increasing $\bar{\sigma}_{H'-X}$ above the region of minimal mixing (and into the region labeled with that name in Fig.~\ref{fig:minmax}) makes oscillations less effective. 

The curves in Fig.~\ref{fig:gosc} peak at $\Delta V(z) 
\sim \delta$, transitioning from a mixing regime in which the medium suppresses $H$--$H'$ conversion (high redshift) to a regime in which the conversion rate is independent of $\delta$ (low redshift). From Eqs.~\eqref{gosc}, \eqref{gammaheq}, and \eqref{vheq}, we see that $\Gamma_\textrm{osc}$ scales like $z^{-4}$ on the high-$z$ side of the peak, being suppressed not only by $\sin^2 2\theta_m$ but by the quantum Zeno effect, \textit{i.e.}, the prevention of quantum coherence development by rapid scattering. On the low-$z$ side of the peak, $\Gamma_\textrm{osc}$ drops off like $z^4$.

For the hydrogen portal to be consistent with cosmology, it must also be in equilibrium as the universe reionizes. Rather than modeling the radiation field as it evolves over the epochs of cosmic dawn and reionization, we choose $z=7$ as a representative redshift and assume that there are $\mathcal{O}(10)$ ionizing photons per baryon at this time. While the influence of photons on $\Delta V$ continues to be marginal, photoionization greatly enhances $\Gamma_c$ because of its large cross section $\sigma_I \sim 10^7 \sigma_T$ near threshold. ($\sigma_T$ is the Thomson cross section.)

\begin{figure}
    \centering
    \includegraphics[width=0.45\textwidth]{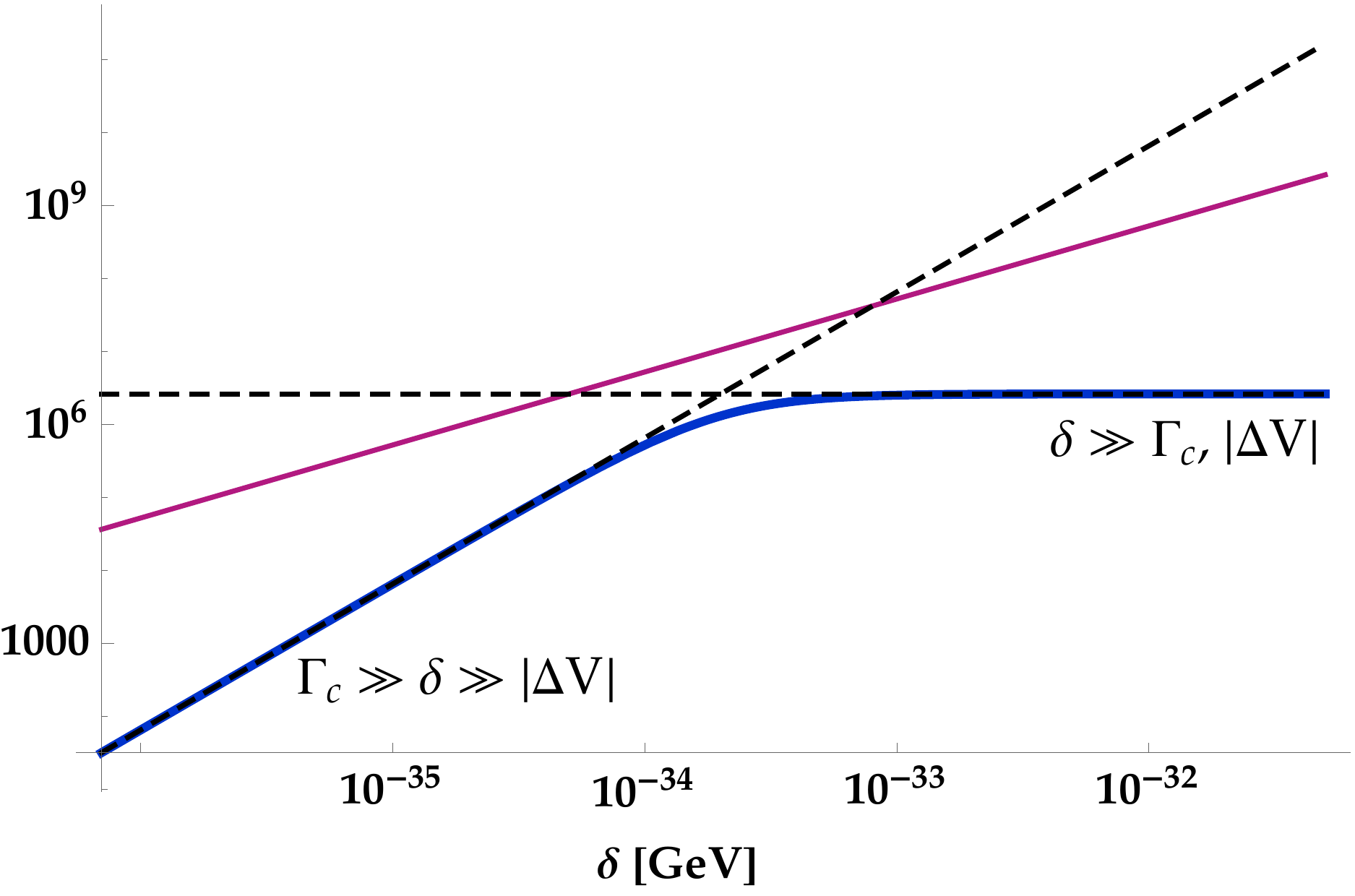}
    \caption{Comparison of rates during the epoch of reionization: $\Gamma_\textrm{osc} / H$ (blue) and $\delta / H$ (purple) evaluated at $z = 7$. Dashed lines show the limiting behaviors. At large values of $\delta$, $\Gamma_\textrm{osc} \sim \Gamma_c / 4$. At smaller values, $\Gamma_\textrm{osc} \sim \delta^2 / \Gamma_c$. Within the parameter space of Fig.~\ref{fig:minmax}, the model is in the quantum Zeno regime, but mixing remains efficient on the Hubble timescale.}
    \label{fig:reion}
\end{figure}

Figure~\ref{fig:reion} displays the ratios $\Gamma_\textrm{osc} / H$ and $\delta / H$ at reionization ($z=7$) as functions of $\delta$. (In calculating $\Gamma_\textrm{osc}$, we implicitly adopt a cosmology in which intergalactic magnetic fields originate from astrophysical processes and do not strongly affect $H$--$H'$ mixing until after reionization.) Mixing is suppressed by the quantum Zeno effect for $\delta \lesssim \Gamma_c$, thus implying a minimum $\delta$ such that $\Gamma_\textrm{osc} \gtrsim H$ at reionization. That cutoff is shown in Fig.~\ref{fig:minmax} as the border of the region marked \textit{Reion.} It happens to be a factor of $\sim 25$ above the lower limit that comes from requiring that mixing equilibrate before $z \sim 20$.

With $H$ and $H'$ rapidly interconverting, thermal equilibration of $H'$ and $X$ at best brings the gas temperature down to
\begin{equation}
T_g \sim T_g^0 \frac{n_H + n_{H'}}{n_H + n_{H'} + n_X} \sim T_g^0 \frac{1}{1 + \frac{6~\textrm{GeV}}{m_X}}, \label{maxcooling}
\end{equation}
relative to the gas temperature $T_g^0$ in the standard scenario of adiabatic cooling \cite{Barkana:2018lgd}. The thermal energy of DM has been neglected in the rightmost expression in Eq.~\eqref{maxcooling}.  Realistically, $T_g$ cannot quite reach this idealized limit because the bulk relative velocities of the gas and DM are dissipated into heat by $H'$--$X$ scattering \cite{Dvorkin:2013cea,Munoz:2015bca}. The approximation is adequate for our purposes, however. Matching to the EDGES anomaly thus implies $m_X \sim 2$~GeV in this simple case.

With the momentum-transfer cross section parametrized as
\begin{equation}
\bar{\sigma}_{H'-X} = \int d\cos\theta \left( 1 - \cos\theta \right) \frac{d \sigma_{H'-X}}{d \cos\theta} = \sigma_0 | \vec{v}_m |^n
\end{equation}
in terms of the relative velocity $\vec{v}_m$, the heating rate of the gas becomes \cite{Dvorkin:2013cea, Munoz:2015bca, Munoz:2017qpy}
\begin{equation}
\dot{Q}_g \sim - \frac{\rho_X \sigma_0 m_H}{\left( m_H + m_X \right)^2} \left( \frac{T_g}{m_H} \right)^{\frac{n+1}{2}} \frac{2^{\frac{5+n}{2}} \Gamma \left( 3 + \frac{n}{2} \right)}{\sqrt{\pi}} T_g. \label{qgdot}
\end{equation}
Eq.~\eqref{maxcooling} is valid provided that $| \dot{Q}_g | H^{-1}$ is at least on the order of the gas thermal energy during a period in which $H$--$H'$ mixing is in equilibrium. This condition is required so that thermal equilibrium may be attained. 

The lower bounds on $\bar{\sigma}_{H'-X}$ shown in Fig.~\ref{fig:minmax} derive from this condition. Two benchmark cases are shown: $n=-4$ and $n=0$. The first of these is comparable to hydrogen-cooling models where SM--DM scattering is facilitated by a light mediator. In those models, the cross section needs to exhibit a steep velocity-dependence in order to avoid frequent interaction during the recombination era \cite{Berlin:2018sjs,Barkana:2018qrx,Slatyer:2018aqg,Fraser:2018acy,Kovetz:2018zan,Liu:2019knx}. The second of the two cases illustrates how that constraint on scattering can effectively be exchanged, in the hydrogen portal scenario, for a constraint on mixing (\textit{i.e.}, not allowing mixing to equilibrate before $z \sim 200$). This accounts for the upper limits on $\delta$ shown in Fig.~\ref{fig:minmax}.

No such constraint applies to $n = -4$ when $\bar{\sigma}_{H'-X}$ is less than $\sim 4 \%$ of $\sigma_{H-H}$. At these relatively weaker cross sections, heat transfer between $H'$ and $X$ is inefficient prior to $z \sim 200$. Up to a point, no harm is done by having mixing come into equilibrium earlier than this, hence the swath in Fig.~\ref{fig:minmax} that cuts through the two regions labeled \textit{CMB}. Values of $\delta$ far above the range plotted in Fig.~\ref{fig:minmax} may eventually affect recombination enough to be ruled out, but we do not extend the analysis up to this regime. Larger $\delta$ implies a lower natural scale for UV physics generating the mixing from Eq.~\eqref{lambda}, but the sensitivity is mild as $\text{d} \log \Lambda / \text{d} \log \delta = -1/8$.

Based on the analysis above, we conclude that the hydrogen portal is a viable way to cool $T_g$ to the temperature inferred from the EDGES absorption trough. It is worth emphasizing that our analysis has included many simplifying assumptions to focus on the simplest region of parameter space to study, and it would be interesting to relax these in a variety of directions. In particular, the focus here has been on thermal DM scenarios in which $H'$--$X$ heat transfer is complete. This restriction implies DM mass near the GeV scale, but lighter masses are possible as well if these assumptions are relaxed.

Hydrogen mixing raises a number of issues that we have not yet addressed. On the high energy side, it remains to establish what underlying physics might give rise to the mixing---or even if doing so is plausible at all. Unlike neutrinos or photons, hydrogen atoms are composite particles. Coupling $H$ to a new state $H'$ seems contrived unless the latter knows about the SM structure in some way.

Introducing a mirror sector that parallels the SM particle content is a well-studied model building framework in which to address this concern.  Ref.~\cite{companion} shows that a mirror model with an effective dimension-12 partonic mixing operator implements the mechanism discussed herein with its features symmetry-protected. Furthermore it establishes leptoquarks as a plausible origin of the mixing operator in a toy UV completion. As shown in that work, the threat of proton decay can be easily avoided despite the violation of baryon number.

On the cosmological side, having half the baryon density during some stretch of cosmic history is an eyebrow-raising modification. There are bound to be effects on star formation, likely with other consequences for 21 cm cosmology. Detailed study is warranted here.

One reasonable prediction is that some mirror hydrogen clouds are destined for cooling and collapse, possibly resulting in black holes and mirror stars \cite{companion}. Regardless of outcome, some fraction of the baryon density almost certainly finds itself stuck in the mirror sector even after reionization. In other words, the hydrogen portal \textit{predicts} missing baryons in the low-$z$ universe, and to our knowledge is the first model to do so.

21 centimeter cosmology has truly opened a new window to the universe, and promises to illuminate for us the heretofore-dark ages. Not only will we learn about the formation of large scale structure and the timeline of reionization, but with some luck the 21 cm sky may also provide a surprise glimpse of new fundamental physics. Forthcoming experiments will test the EDGES result and furnish further insight, and we may look forward to seeing the fruits of outstanding observational efforts come to bear in the coming years. 

We thank Samuel Alipour-fard, Guido D'Amico, Robert McGehee, Paolo Panci, and Yiming Zhong for comments on a draft of this manuscript. LJ thanks Anna Schauer for insights into first-star formation and for suggesting a connection to direct-collapse black holes. SK thanks Vera Gluscevic for presenting an enlightening seminar on 21cm cosmology at the KITP in December 2019.

The work of LJ was supported by NSF Grant No. PHY-1914242 and by NASA through the NASA Hubble Fellowship grant \# HST-HF2-51461.001-A awarded by the Space Telescope Science Institute, which is operated by the Association of Universities for Research in Astronomy, Incorporated, under NASA contract NAS5-26555. The work of SK was supported in part by the Department of Energy under the grant DE-SC0250757, and by a Mafalda and Reinhard Oehme Postdoctoral Fellowship from the Enrico Fermi Institute at the University of Chicago.

\bibliography{hydrogen}% Produces the bibliography via BibTeX.

\end{document}